\documentclass[aps,prl,showpacs,notitlepage,reprint,floatfix]{revtex4-1}
\usepackage{graphicx}
\usepackage{dcolumn}
\usepackage{amsmath}
\usepackage{bm}
\usepackage{todonotes}
\usepackage[utf8]{inputenc}
\usepackage{amsmath}
\usepackage{natbib}
\usepackage{graphicx}
\usepackage{caption}
\usepackage{subcaption}
\usepackage{lipsum}
\usepackage{float}
\newcommand{\bn}[1]{\mbox{\boldmath$#1$}}

\newcommand{\beq}{\begin{equation}}
\newcommand{\eeq}{\end {equation}}
\newcommand{\bea}{\begin{eqnarray}}
\newcommand{\eea}{\end{eqnarray}}

\begin{document}

\title{Magneto-optical polarisation texturing}
\author{{\rm {K. Koksal}}$^{1,2}$, F. Tambag$^{1}$, J. Berakdar$^{2}$, M. Babiker$^{3}$ } 
\affiliation{$^{1}$Physics Department, Bitlis Eren University, Bitlis, Turkey}
\affiliation{$^{2}$Institut für Physik, Martin-Luther Universität Halle-Wittenberg, D-$06099$ Halle, Germany}
\affiliation{$^{3}$School of Physics, Engineering and Technology, University of York, York, YO$10$ $5$DD, UK} 
\date{\today}

\begin{abstract}
Left  and right circularly polarized transverse electromagnetic waves propagate at slightly different speeds in a magnetic material leading to a polarization rotation by an amount proportional to the projection of the magnetic field along the direction of the wave propagation.
We show how this magneto-optical effect can serve as a vectorial polarization shaper if the input mode is either a radially-polarised or an azimuthally-polarised Laguerre-Gaussian (LG) mode.  The specific polarization map of the output field can be achieved by choosing  appropriately the magnetic material and/or its geometry. We show further that when the LG beam waist is comparable to the wavelength (i.e. $w_0\approx \lambda$) the fields are no longer purely transverse but acquire an additional longitudinal (axial) component.  We demonstrate how this modifies the polarisation texturing. 
\end{abstract}


\maketitle

  \section{Introduction}
A number of gyromagnetic phenomena occur as  light travels at different speeds in a magneto-optic crystal, depending on its state of polarisation, resulting in  a rotation of the polarisation vector upon traversing the sample. A prominent example is the  Faraday effect which is central to numerous  applications such as  magnetic field measurements \cite{vle2019},  Faraday isolators \cite{davut2019enhanced} operating at different wavelengths \cite{vojna2020}, and others \cite{ghaderi2021,schatz1969faraday,bennett1965faraday,serber1932theory,boswarva1962faraday}. 

In a conventional magneto-optical set-up, e.g. in a Faraday-rotation experiment, a linearly-polarised light passing through a length $z$ of the magneto-optic crystal experiences a rotation in the direction of its polarization by an angle $\Theta$ whose magnitude depends on the Verdet constant of the material \cite{vojna2019verdet,vojna2020faraday}, the length of the sample and the value of the axial magnetic field $B$.

Consider now this conventional   set-up, but instead of the (moderate intensity) linearly-polarised light input mode we have  {\underline {either}}
a radially-polarised {\underline {or}} an azimuthally-polarised Laguerre-Gaussian (LG) mode\cite{allen1999,andrews2012,zhan2009,Rosales-Guzmán_2018,koksal2023}. What mode polarization distribution  would we have as the mode travels through the magnetic sample?
Here,  we first develop the analysis  for  an incident  purely radially-polarised mode. The case of an incident purely azimuthally-polarised mode can be treated along similar lines.  A recent interesting report evaluated the properties of optical fields formed as a combination of radially and azimuthally polarised modes \cite{herrero2024}, but did not consider the role of the Faraday effect in that context.

\section{Radially-polarised input mode}

We first focus in some detail on the case of a cylindrical radially-polarised LG mode in the  paraxial approximation with a waist  $w_0$ at focus. The transverse component of the electric field vector of this mode is along the radial unit vector ${\bn {\hat {\rho}}}$ but it is well-known that this is expressible in terms of left and right-circularly polarized LG modes.  Inside the medium the left and the right circularly-polarised field components propagate at different speeds due refractive indices $n^{(\pm)}$.  In general the fields include a longitudinal (axial) component along the propagation direction ${\bn {\hat z}}$ which comes into play when the beam width $w_0$ is comparable to the wavelength $\lambda$. The total field is thus as follows


\begin{widetext}
\begin{equation}
{\bf E}=\frac{1}{\sqrt{2}}\left[ick_z({\bn {\hat x}}-i{\bn {\hat y}}){\cal F}^{(1)}-{\bn {\hat z}}c\left\{\frac{\partial {\cal F}^{(1)}}{\partial x}-i\frac{\partial {\cal F}^{(1)}}{\partial y}\right\}\right]e^{i n^{(-)} k_zz}
+\left[ick_z({\bn {\hat x}}+i{\bn {\hat y}}){\cal F}^{(2)}-{\bn {\hat z}}c\left\{\frac{\partial {\cal F}^{(2)}}{\partial x}+i\frac{\partial {\cal F}^{(2)}}{\partial y}\right\}\right]e^{i n^{(+)} k_zz}
\label{emfields1}
\end{equation} 
\end{widetext}

\beq{\cal F}^{(1)}({\bf r})= e^{i\phi}u(\rho,z);\;\;\;
{\cal F}^{(2)}({\bf r})= e^{-i\phi}u(\rho,z)
\label{eff12}
\eeq 

Here a caret denotes a unit vector and $u(\rho)$ is the amplitude function of the  LG mode of winding number $\ell=1$ which we assume is a cylindrical doughnut mode (radial number $p=0$) \cite{zhan2009}.  We have 
\beq
u(\rho)=A_0 e^{-\frac{\rho^2}{w_0^2}}  \left(\frac{\sqrt{2} \rho}{w_0}\right)L^{1}_0\left(\frac{2\rho^2}{w_0^2}\right).
\label{LG}
\eeq
where $A_0$ is a normalisation constant, determined in terms of the applied power ${\cal P}$ of the mode and is such that \cite{babiker2024}
\beq
 {A}_0^2=\frac{4{\cal P}}{\pi\omega^2n_0^2\epsilon_0 cw_0^2}\label{Azero}
\eeq

$n_0$ is the refractive index of the material in the absence of the magnetic field, $L_p^{|\ell|}$ with $ \ell=1, p=0$, is the associated Laguerre polynomial, $k_z$ is the axial wavevector, $n^{(\pm)}$ are the refractive indices of the magneto-optic  medium for left-hand circularly-polarised light $(+)$ and right-hand circularly-polarised light $(-)$, respectively. 
Within the magneto-optic medium i.e. in the presence of a magnetic field, the change in the refractive indices is responsible for the Faraday rotation such that \cite{berman2010optical}

 \beq
 n^{(-)}- n^{(+)}=\gamma\Theta
 \label{gammatheta}
 \eeq
 where $\Theta$ is the rotation angle given by the standard expression
 \beq
\Theta(z)={\cal V}Bz\label{Thetaz}
 \eeq
 where ${\cal V}$ is the Verdet constant of the material,  B is the applied axial magnetic field and $z$ is the axial position in the medium.   For a medium of length $L$ the parameter $\gamma$ is such that
 \beq
 \gamma (L) =\frac{2 c}{\omega L}\label{gammaell}.
 \eeq
Clearly the difference in the refractive indices  $n^{(-)}- n^{(+)}$ is a property of the medium and can be seen from Eq.(\ref{gammatheta}), Eq.(\ref{Thetaz}) (putting $z=L$)  and (\ref{gammaell}) to be independent of $L$.  The left-hand circular polarisation and the right hand one produce equal but opposite changes (in the sign) of the rotation angle so that Eq.(\ref{gammatheta}) suggests the following forms of the individual refractive indices
\beq
n^{(-)}=n_0+\left(\frac{\gamma\Theta}{2}\right);\;\;\;\;\;n^{(+)}=n_0-\left(\frac{\gamma\Theta}{2}\right)\label{nplusminus}
\eeq
where, as pointed out above, $n_0$ is the refractive index in the absence of the magnetic field.  
In general we are interested in the propagation along the z axis for unspecified axial position $z$, in which case we replace $L$ by z
to obtain the fields at a general point ${\bf r}=(\rho,\phi,z)$ inside the medium, so that Eq.(\ref{gammaell})  can be written as  $z\gamma(z) =2 c/\omega$. In general, the propagating fields have both transverse and longitudinal components, so it is convenient to deal with these in turn.

\section{Transverse fields}

The transverse fields are those involving the unit vectors ${\bn {\hat x}}$ and ${\bn {\hat y}}$ which can be deduced from 
Eq.(\ref{emfields1}) with Eq.(\ref{eff12}). 
Substituting for $n^{(-)}$ and $n^{(+)}$ from Eq.(\ref{nplusminus}), we have 
$k_zz\gamma\Theta/2=\Theta$. The transverse electric field at ${\bf r}=({\bn {\rho}},\phi, z)$
in the material can then be written as follows
\begin{widetext}
\begin{equation}
{\bf{E}}_{T}(\rho,\phi,z)=ick_z \frac{u(\rho)}{\sqrt{2}}\left\{{\bn {\hat{x}}}\left[e^{(i k_z z n_0 - i\Theta(z)+i\phi)}+e^{(i k_z z n_0 + i\Theta(z)-i\phi)} \right]
- i{\bn {\hat{y}}}\left[e^{(i k_z zn_0 -i\Theta(z)+i\phi)}-e^{(i k_z zn_0 + i\Theta(z)-i\phi)} \right]\right\} 
\label{eq:refrac3}
\end{equation}
\end{widetext}
%
This is  the electric field inside the medium in the absence of the longitudinal component.  It is thus equal to the total field for beam widths greater than the wavelength (i.e. $w_0\gg\lambda$) when the longitudinal component is negligible.  The expression for ${\bf{E}}_{T}$ simplifies further to obtain 
\begin{equation}
 {\bf{E}}_T=\sqrt{2}ick_zu(\rho)e^{i k_z zn_0}\left\{\cos(\phi-\Theta) {\bn {\hat{x}}}+\sin(\phi-\Theta){\bn {\hat{y}}}\right\}  
 \label{eq:radelfield1}
\end{equation}

This is the transverse electric vector field of the light beam after traversing the length $z$ from the focal plane in the magneto-optic medium under the applied magnetic field. The  corresponding intensity distribution is
\bea
 I_T(\rho,\phi,z)&=&\frac{1}{2}\epsilon_0cn_0^2{\bf{E}}_T\cdot {\bf{E}^*}_T\nonumber\\
 &=&\epsilon_0 c^3k_z^2 n_0^2u^2(\rho)\left\{\cos^2[\phi-\Theta] +\sin^2[\phi-\Theta]\right\}\nonumber\\
 &=& \epsilon_0 c^3k_z^2 n_0^2 u^2(\rho)
 \label{eq:radelfield}
\eea
This is independent of $\phi$ and $\Theta$. However, in view of Eq.(\ref{eq:radelfield1}) there exists a transverse polarisation vector ${\bn {\hat {\epsilon}}}_T$, which is a unit vector in  the direction of the transverse electric field.  It is 
\beq
{\bn {\hat {\epsilon}}}_T(\phi,\Theta)=\cos(\phi-\Theta) {\bn {\hat{x}}}+\sin(\phi-\Theta){\bn {\hat{y}}}
\label{textu}
\eeq
We note first that in the absence of the axial  magnetic field ($B=0$) the rotation angle $\Theta$ is zero, we then have
\beq
{\bn {\hat {\epsilon}}}(\phi,0)=\cos(\phi) {\bn {\hat{x}}}+\sin(\phi){\bn {\hat{y}}}={\bn {\hat {\rho}}}\label{epsil}
\eeq
which confirms that the input mode is radially-polarised.  However, for a given material with a known Verdet constant ${\cal V}$ and a fixed length $L$, the Faraday rotation angle $\Theta$ can be modified by a change in the applied axial magnetic field $B$. For a demonstration, we consider  Terbium Gallium Garnet (TGG) as the magneto-optic material, which is an optical isolator with small losses. Note that TGG has a strong frequency dependent Verdet constant. Its Verdet constant at $\lambda=632$ nm is   $-131$  rad/(T·m) \cite{Majeed:13,app9153160}. We choose the 
thickness  as $L=0.1$ m.  So, in order to achieve a Faraday rotation angle $\Theta=\pi/2$, the applied magnetic field should be
\beq
B=\frac{\Theta}{{\cal V}L}\approx 0.12 {\rm {T}}.
\eeq
 On the other hand, there exists a variety of materials that can serve our purpose and so we can consider other magneto-optical materials with other frequency dependence \cite{app9153160}. 
 Figure 1 displays nine sub-figures, each showing the light intensity profile superimposed on which is a set of arrows indicating the direction of the local unit vector ${\bn {\hat {\epsilon}}}(\phi,\Theta)$. Each sub-figure is for a specific value of $\Theta$ and the set  spans the value range $(\Theta=0\; {\rm {to}}\; 2\pi)$.  This figure demonstrates the control of polarisation using the Faraday effect.  It shows the nature of the changes in the wave polarisation as the angle $\Theta$ varies by changing the magnitude of the magnetic field. In the first sub-figure $\Theta=0$ the arrows indicate a radially-polarised mode, but as $\Theta$ increases the distribution of arrows indicates azimuthal polarisation for $\Theta=\pi/2$.  It is straightforward to see that Eq.(\ref{textu}) yields on substituting $\Theta=\pi/2$ the following expression
\beq
{\bn {\hat {\epsilon}}}_T(\phi,\pi/2)=-\sin(\phi) {\bn {\hat{x}}}+\cos(\phi){\bn {\hat{y}}}={\bn {\hat {\phi}}}\label{epsil2},
\eeq
which confirms that for this $\Theta$ the original radially-polarised mode has become an azimuthally-polarised mode. At $\Theta=\pi$ we obtain a radially-polarised mode in which the arrows point inwards everywhere and at $\Theta=3\pi/2$ we obtain an azimuthally-polarised mode in which the arrows point in the opposite direction to the one at $\Theta=\pi/2$.  Finally at $\Theta=2\pi$ we are back to the radial polarisation coinciding with the case for which $\Theta=0$.
\begin{figure*}[ht!]
      \centering
     \includegraphics[width=\textwidth]{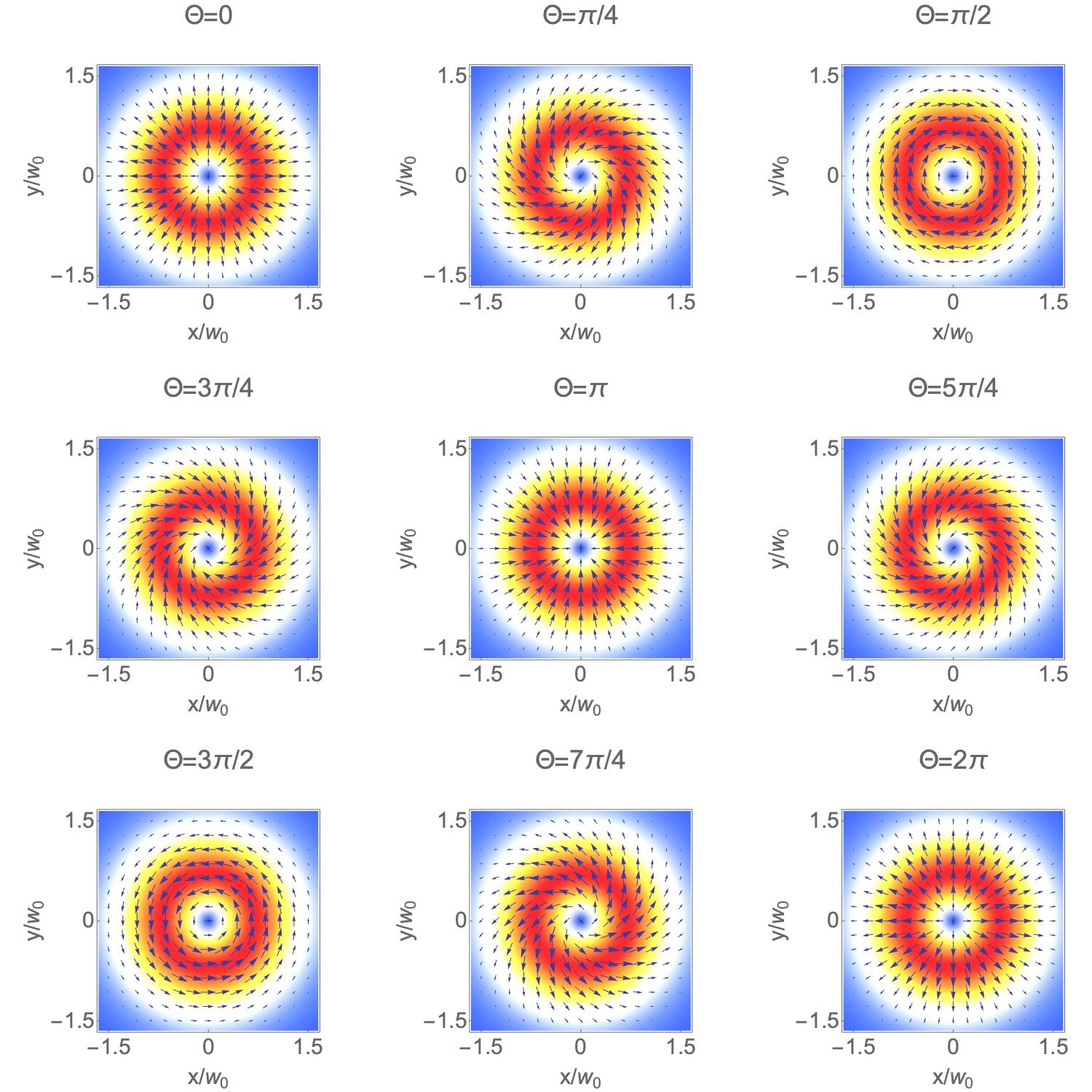}
      \caption{The doughnut light intensity profiles superimposed on which are sets of arrows indicating the direction of the local unit vector ${\bn {\hat {\epsilon}}}(\phi,\Theta)$. Each sub-figure is for a specific value of $\Theta$ and the set of sub-figures spans the value range $(\Theta=0\; {\rm {to}}\; 2\pi)$, with the polarisation changing from the initial outwards radial polarisation $\Theta=0$ to azimuthal ($\pi/2$) to radial in opposite directions ($\pi$), then azimuthal in apposite direction ($3\pi/2$), then radial in the original outwards direction ($2\pi$)}
      \label{fig:figl1full}
  \end{figure*}
%
\\
A visual means of detecting the angle of rotation can be realised by passing the beam emerging from the magneto-optic medium of length $L$ through a linear polariser oriented along the  $x$ direction, so that the electric field, referred to as ${\bf E}_{Tx}$, is given by the first term of Eq.(\ref{eq:radelfield1}) and the intensity, referred to as $I_{Tx}$, by the first term of the middle equation in Eq.(\ref{eq:radelfield}).  In general for any axial position $z$ we have
\begin{equation}
\begin{split}
 &{\bf{E}}_{Tx}=\sqrt{2}ick_zu(\rho)e^{i k_z zn_0}\cos(\phi-\Theta) {\bn {\hat{x}}}\\
 &I_{Tx} =\epsilon_0 c^3k_z^2 n_0^2 u^2(\rho)\cos^2(\phi-\Theta)
 \label{eq:radelfield2}
 \end{split}
\end{equation}
so that both the transverse electric field and its intensity are now dependent on both $\phi$ and $\Theta$.\\
Figure 2 displays on the upper panel the arrow distributions representing the polarisation function ${\bn {\hat {\epsilon}}}_T$, Eq.(\ref{epsil}) in the transverse plane for different values of $\Theta$.  This is before the beam enters the linear polariser and in the lower panel are displayed the corresponding intensity profiles after the beam has passed through the linear polariser.  In the sub-figure labeled $\Theta=0$  the intensity gap is vertical but as $\Theta$ increases, it is seen that the angle of the gap gradually changes until it becomes horizontal when $\Theta=\pi/2$, which is the effect of the linear polariser in transforming the beam to an azimuthally-polarised beam. The tilt angle in the intensity gap is a measure of the value of the Faraday rotation $\Theta$, as has been experimentally demonstrated recently \cite{tambag2023}.\\
\begin{figure*}[ht!]
      \centering
     \includegraphics[width=\textwidth]{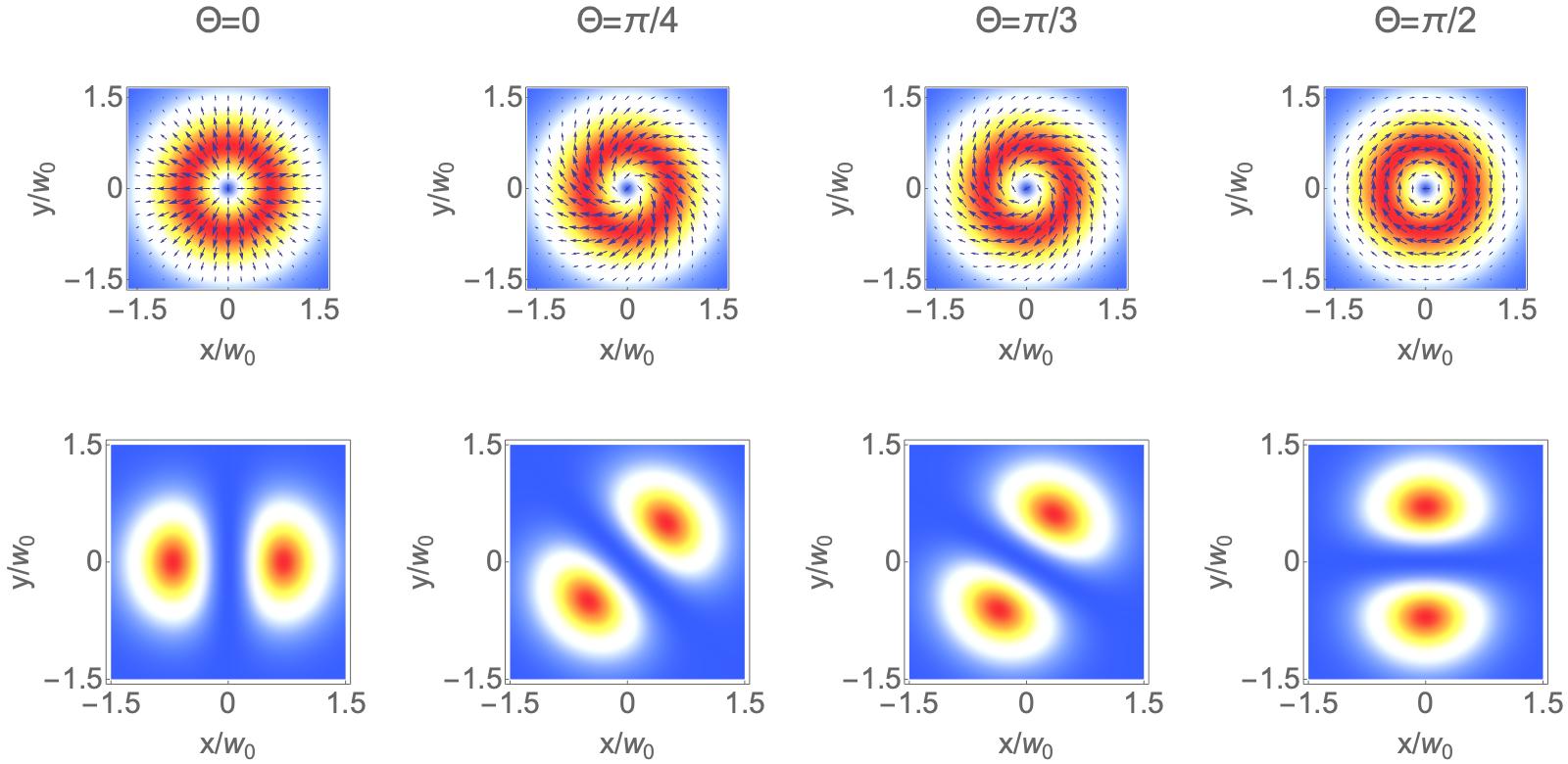}
      \caption{Upper panel: the arrow distributions representing the polarisation vector function in the transverse plane for different values $\Theta$ between $0$ (radial) and $\pi/2$ (azimuthal).  This is before the beam enters the linear polariser.   Lower panel: the corresponding intensity profiles after the beam has passed through the linear polariser.}
      \label{fig:figl1}
  \end{figure*}  
%
\section{Azimuthally-polarised input}
We now turn to consider in less detail the same Faraday set up we have considered above, but now we have an azimuthally polarized LG mode that is made to pass through a magneto-optic crystal subject to an axial magnetic field $B$.  It is straightforward to show that instead of Eq.(\ref{eq:radelfield1}) we now have
\begin{equation}
 {\bf{E}}(\rho,\phi,L)=\sqrt{2}ick_zu(\rho)e^{i k_z zn_0}\left\{\sin(\phi-\Theta) {\bn {\hat{x}}}+\cos(\phi-\Theta){\bn {\hat{y}}}\right\}  
 \label{eq:radelfield3}
\end{equation}
The polarisation vector is then the expression between the brackets in Eq.(\ref{eq:radelfield3}) and we note that the cosine and sine terms are interchanged relative to the polarisation vector in  the radially-polarised case.  We have verified by explicit evaluations that a figure like Fig.1 can be constructed for the azimuthally-polarised case.  This would start at $\Theta=0$ with all arrows pointing in azimuthal direction and as $\Theta$ increases the polarisation vectors change smoothly to radial polarisation at $\Theta=\pi/2$ and as $\Theta$ increases further the polarisation changes again to azimuthally polarised in the opposite direction for $\Theta=\pi$, then to opposite radial at $\Theta=3\pi/2$ and finally to the azimuthally-polarised at $2\pi$.\\
Figure 3 is the azimuthally-polarised analogue of Fig.2.  The Upper panel shows the changes of the polarisation arrow distributions of the azimuthally-polarised mode as $\Theta$ increases from $\Theta=0$ as an azimuthally-polarised mode up to $2\pi$ where the polarisation changes to radial.  The lower panel shows the corresponding intensity after the azimuthally-polarised beam passes through a linear polariser. Once again from the inclination of the intensity gaps in the lower panel we can determine the angle $\Theta$.
\begin{figure*}[ht!]
      \centering
     \includegraphics[width=\textwidth]{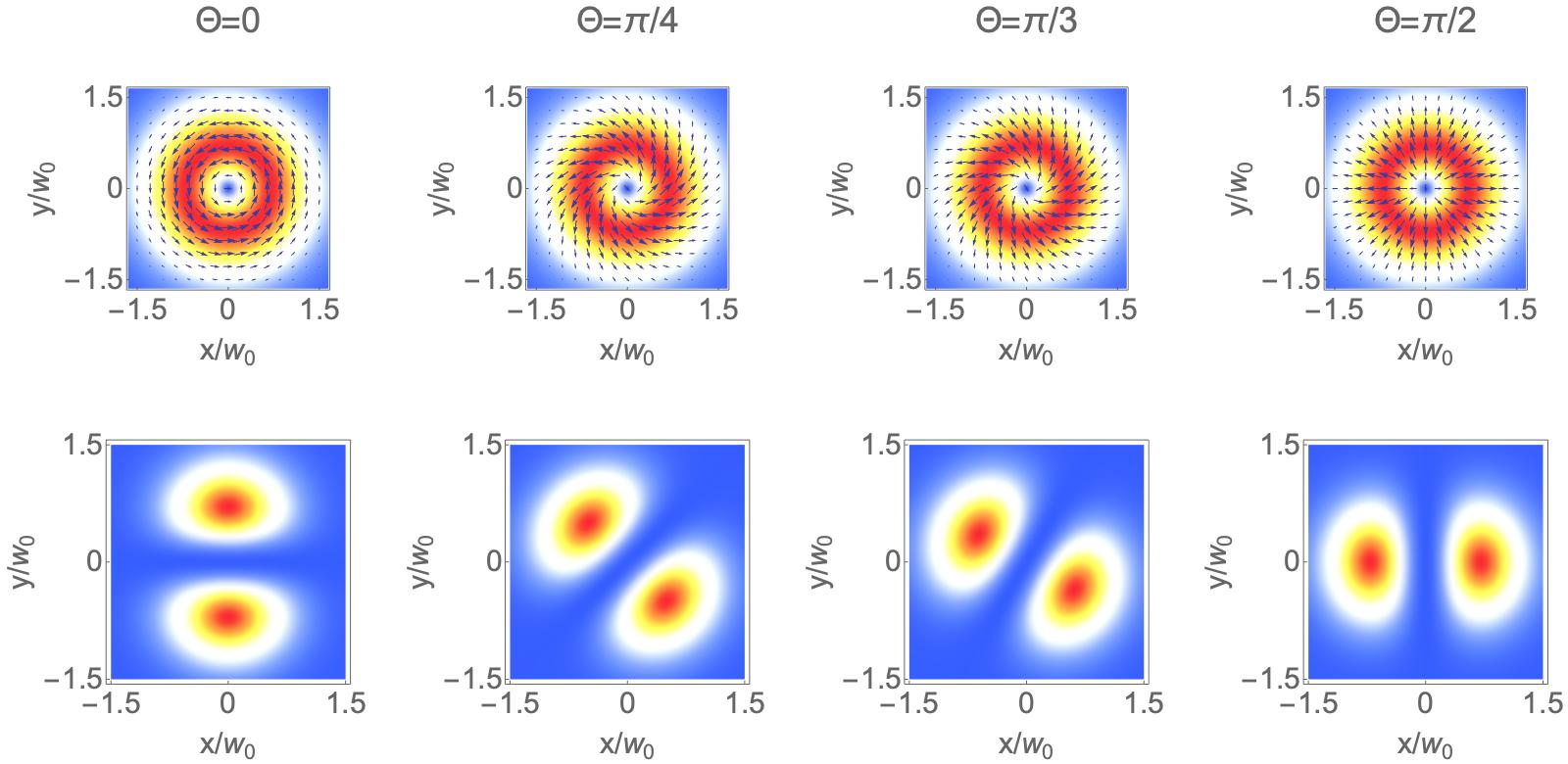}
      \caption{Upper panel: the arrow distributions representing the polarisation vector function in the transverse plane for different values $\Theta$ between $0$ (azimuthal) and $\pi/2$ (radial).  This is before the beam enters the linear polariser.   Lower panel: the corresponding intensity profiles after the beam has passed through the linear polariser.}
      \label{fig:figl1az}
  \end{figure*}

  \section{Longitudinal Field}
As pointed out above, the longitudinal component of the field comes into play when $w_0$ is comparable to the wavelength $\lambda$.  We return to the case of the radially-polarised input beam. The longitudinal field is given by the z-component of the full expression of the electric field vector, as can be deduced from Eq.(\ref{emfields1}). We have
\begin{widetext}
\begin{equation}
 E_z=-c\frac{1}{\sqrt{2}}\left(\left\{\frac{\partial {\cal F}^{(1)}}{\partial x}-i\frac{\partial {\cal F}^{(1)}}{\partial y}\right\}e^{i n^{(-)} k_zz}+\left\{\frac{\partial {\cal F}^{(2)}}{\partial x}+i\frac{\partial {\cal F}^{(2)}}{\partial y}\right\}e^{i n^{(+)} k_zz}\right)
\label{emfieldsz}
\end{equation} 
\end{widetext}
where ${\cal F}^{(1,2)}$ are given by Eqs.(\ref{eff12}).
The Cartesian derivatives are straightforward to evaluate. We find
\beq 
\frac{\partial {\cal F}^{(1)}}{\partial x}=\left\{\cos\phi\;{u'(\rho)}-i\frac{1}{\rho}\sin\phi \;{u(\rho)}\right\}e^{i\phi}\label{dfx}
\eeq
\beq 
\frac{\partial {\cal F}^{(1)}}{\partial y}=\left\{\sin\phi\;{u'(\rho)}+i\frac{1}{\rho}\cos\phi \;{u(\rho)}\right\}e^{i\phi}\label{dfy}
\eeq
\beq 
\frac{\partial {\cal F}^{(2)}}{\partial x}=\left\{\cos\phi\;{u'(\rho)}+i\frac{1}{\rho}\sin\phi \;{u(\rho)}\right\}e^{-i\phi}\label{dfx21}
\eeq
and 
\beq 
\frac{\partial {\cal F}^{(2)}}{\partial y}=\left\{\sin\phi\;{u'(\rho)}-i\frac{1}{\rho}\cos\phi\; {u(\rho)}\right\}e^{-i\phi}\label{dfy22}
\eeq
where $u'=\partial u/\partial \rho$.  We have on substituting for ${\cal F}^{(1)}$ using the first equation in  Eq.(\ref{eff12})
\beq
\left\{\frac{\partial {\cal F}^{(1)}}{\partial x}-i\frac{\partial {\cal F}^{(1)}}{\partial y}\right\}e^{i n^{(-)} k_zz}=\left(\frac{\partial u}{\partial \rho}+\frac{u}{\rho}\right)e^{i n^{(-)} k_zz}\label{eez1}
\eeq
Similarly for ${\cal F}^{(2)}$ we obtain
\beq
\left\{\frac{\partial {\cal F}^{(2)}}{\partial x}+i\frac{\partial {\cal F}^{(2)}}{\partial y}\right\}e^{i n^{(+)} k_zz}=\left(\frac{\partial u}{\partial \rho}+\frac{u}{\rho}\right)e^{i n^{(+)} k_zz}\label{eez2}
\eeq
Thus on substituting from Eqs. (\ref{eez1}) and (\ref{eez2}) we find  
\beq
E_z=-\frac{c}{\sqrt{2}}\left(\frac{\partial u}{\partial \rho}+\frac{u}{\rho}\right)\left[e^{in^{(-)}k_zz}+e^{in^{(+)}k_zz}\right]
\eeq
Substituting $n^{(\mp)}=n_0\pm\frac{\gamma\Theta}{2}$ and $k_zz\gamma\Theta/2=\Theta$, we obtain finally
\beq
{\bf E}_z={\bn {\hat{z}}}E_z=-\sqrt{2}c\left(\frac{\partial u}{\partial \rho}+\frac{u}{\rho}\right)\cos(\Theta(z))e^{ik_zn_0z}{\bn {\hat {z}}}
\eeq\\
The total field is the vector sum
\beq
{\bf E}(\rho,\phi,z)={\bf E}_T(\rho,\phi,z)+{\bf E}_z(\rho,\phi,z)
\eeq
The light intensity associated with the electric field is denoted $I_z$ which is
\beq
I_z=\frac{1}{2}\epsilon_0n_0^2c|{\bf E}_z|^2=\epsilon_0n_0^2c^3\left|\frac{\partial u}{\partial \rho}+\frac{u}{\rho}\right|^2\cos^2(\Theta)
\eeq
We find on substituting for $u(\rho)$, Eq.(\ref{LG}) and after some algebra involving evaluation of the derivative $\partial u/\partial\rho$
\beq
I_z({\tilde \rho})=\frac{4}{k_z^2w_0^2}\left(-2+\frac{1}{{\tilde \rho}^2}+{\tilde \rho}^2\right)\cos^2{(\Theta)}I_T({\tilde \rho})\label{eyeza}
\eeq
where ${\tilde \rho}=\rho/w_0$.  Explicitly we have on inserting $I_T({\tilde \rho})$ from Eq.(\ref{eq:radelfield}) and substituting for $A_0$ from Eq.(\ref{Azero})
\beq
I_z({\tilde \rho})=\frac{1}{k_z^2w_0^2}\left(\frac {16 {\cal P}}{\pi w_0^2}\right)\left[1-{\tilde \rho}^2\right]^2\cos^2{(\Theta)}e^{-2{\tilde \rho}^2}\label{eyez}
\eeq
We may then compare $I_z$ with $I_T$ for the same set of parameters. The transverse intensity $I_T$ is given by Eq.(\ref{eq:radelfield2}). This can be written in terms of the same notation after substituting for $u(\rho)$ along with $A_0$
\beq
I_T=\left(\frac{8{\cal P}}{\pi w_0^2}\right){\tilde \rho}^2 e^{-2{\tilde \rho}^2}\label{eyet}
\eeq

\begin{figure}[!ht]\label{Figure 4}
  \centering
\begin{subfigure}[]{0.45\textwidth}
  \centering
  \includegraphics[width=\textwidth]{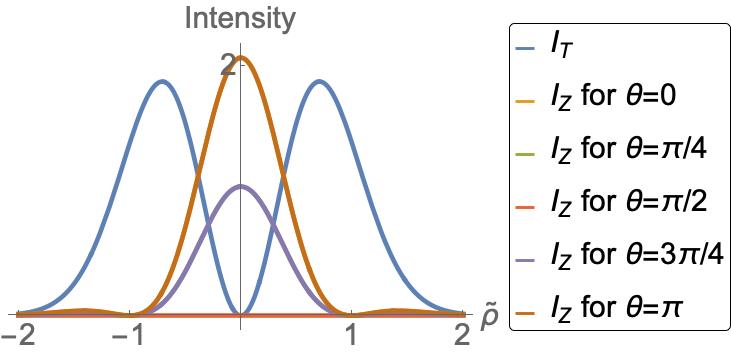}  
  \caption{\small{Variations  with the scaled radial coordinate ${\tilde \rho}=\rho/w_0$ of the longitudinal intensity $I_z$, Eq.(\ref{eyez}) for the same parameters when compared with the transverse intensity $I_T$, Eq.(\ref{eyet}) (the same arbitrary units) for the same power $\cal P$.  Here we have a small beam waist  $w_0=0.5\lambda$ and the plots are for different Faraday angles $\Theta$, but note that $I_z$ is the same for $\Theta=\pi/4$ and $3\pi/4$. It is also seen that $I_z$ is largest for small values of $\Theta$ and is zero for $\Theta=\pi/2$. The longitudinal field intensity $I_z$ is therefore substantial at this value of $w_0$ and is capable of considerable modifications of the polarisation texturing, as Fig. \ref{fig:Longpart} shows.  Note also that $I_z$ is maximum at the core $\rho=0$, in contrast to $I_T$ which vanishes at $\rho=0.$}}
  \label{fig:4a}
\end{subfigure}
\hfill
\begin{subfigure}[]{0.45\textwidth}
  \centering
  \includegraphics[width=\textwidth]{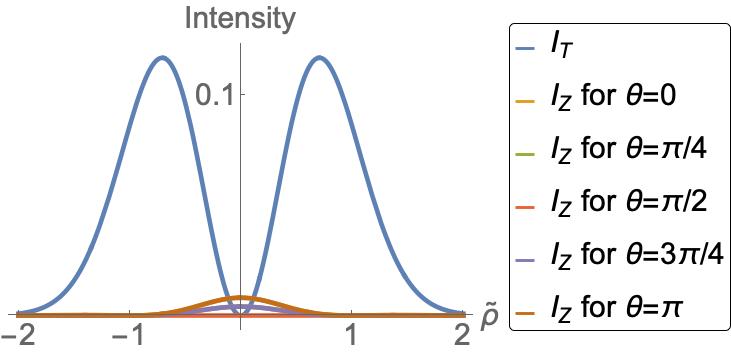}  
  \caption{\small{As in Fig. \ref{fig:4a}, but for the case $w_0=2\lambda$.  Here the longitudinal intensity $I_z$ is much smaller than the transverse intensity $I_T$, suggesting that the polarisation is predominantly due to the transverse components, so, as Fig. 5 confirms, the influence of the longitudinal components on the texturing is small for this larger value of $w_0$.}}
  \label{fig:4b}
\end{subfigure}
\caption{Variations of the transverse and longitudinal field intensities.} 
\label{fig:1Dfig}
\end{figure}

We note the appearance in $I_z$ of the paraxial parameter $\alpha=1/(k_z^2w_0^2)=\frac{\bar {\lambda}^2}{w_0^2}$ where ${\bar \lambda}=\lambda/(2\pi)$.  Figure \ref{fig:4a} displays the two intensities $I_z$  and $I_T$ for  $w_0=0.5\lambda$ for different representative $\Theta$.  We see that the longitudinal field intensity $I_z$ is indeed considerable in magnitude relative to the transverse intensity in this case.  Furthermore, although the transverse mode intensity is zero on the mode axis $\rho=0$, the longitudinal intensity is non-zero. These special features, which become evident when the beam waist is small, are also reflected in the modified polarisation texturing. Figure \ref{fig:4b} concerns the case $w_0=2\lambda$ and we see that the longitudinal intensity is much smaller than the transverse intensity.

\begin{figure*}[ht!]\label{Fig.5}
      \centering
     \includegraphics[width=\textwidth]{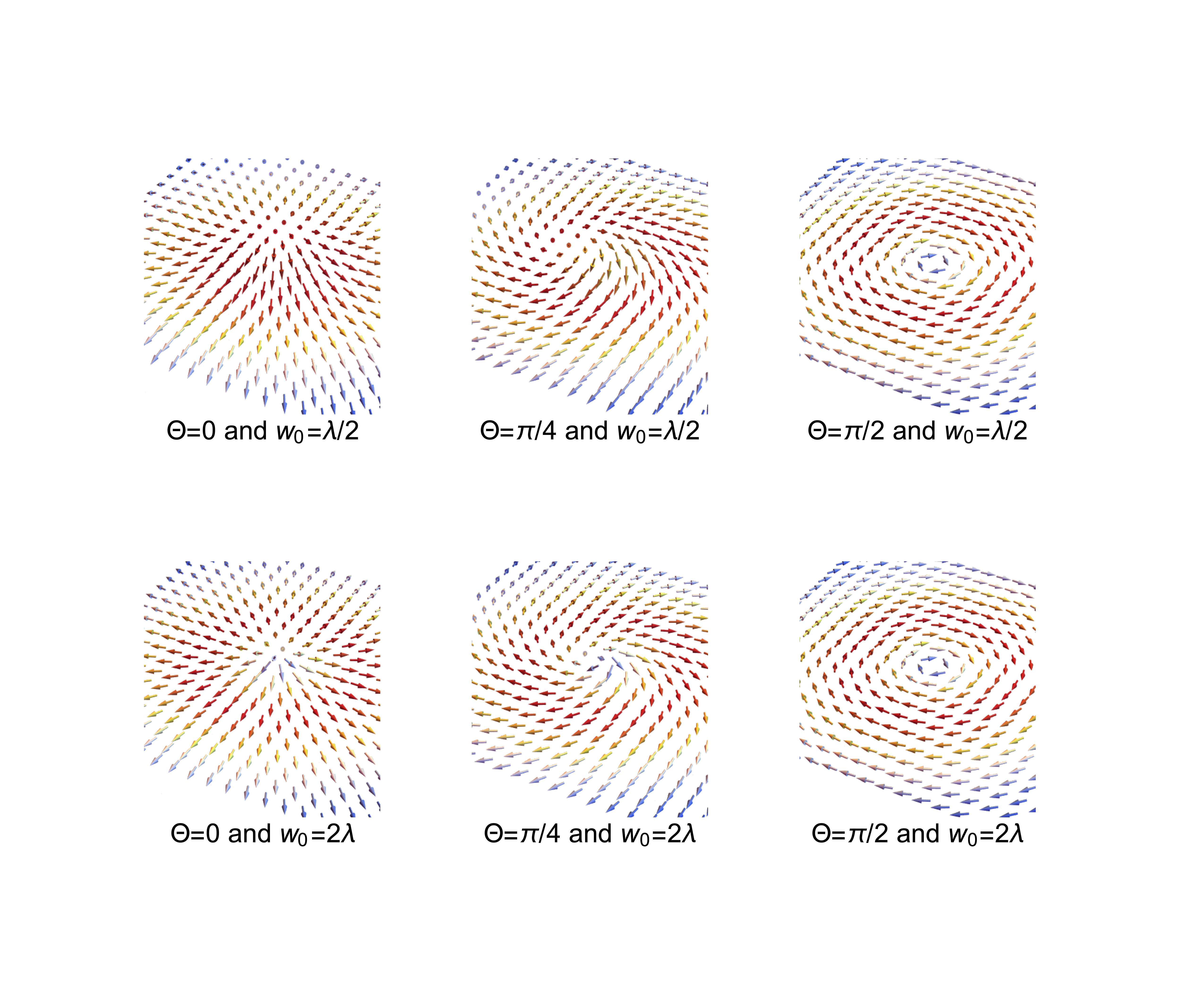}
      \caption{Polarisation textures for different Faraday angle $\Theta$ and in the two cases considered in Fig. 4, namely $w_0=0.5\lambda$ on the top panel and $w_0=2\lambda$ on the bottom panel. Note that the texture plot on the right of the top panel is almost identical to that to the right on the lower panel.  This is because for $\Theta=\pi/2$ the longitudinal field on the top panel is zero and the texture is determined entirely by the transverse component, which for the case of the larger  $w_0=2\lambda$ coincides with the corresponding plot in the lower panel. The color code in this figure is such that the  highest and lowest magnitudes of the arrows correspond, respectively, to the colors red and blue.}
      \label{fig:Longpart}
  \end{figure*}

The results shown in Fig.\ref{fig:4a} indicate that the longitudinal component should have an effect on the texturing and this is indeed so.   This means that the polarisation texturing discussed earlier in the absence of $E_z$ is now changed at every point in both its local direction and magnitude  due to the existence of the longitudinal component.\\ 
Figure \ref{fig:Longpart} compares the polarisation distributions for $w_0=0.5\lambda$ and $w_0=2\lambda$.  We see for the smaller $w_0 (=0.5\lambda) $ the longitudinal field $E_z$ now makes clear contribution in modifying the polarisation texturing, compared with the case $w_0=2\lambda$ where the longitudinal component is small, except for the case where $\Theta=\pi/2$ where the longitudinal component is zero, while for other values of $\Theta$ represented by the first two sets of plots in Fig. \ref{fig:Longpart}, the changes in the textures are clearly evident for small $w_0=0.5\lambda$ case.   Each arrow in Fig. \ref{fig:Longpart} represents both the 3-dimensional direction as well as the magnitude of the total field at that point. 
It is seen that the directions of the arrows are no longer in the plane and they are more out of plane in the case $w_0=0.5\lambda$. Thus the texturing also changes with $\Theta$, or, equivalently, the magnetic field  $B$.

\section{Conclusions}
In conclusion, we have shown  that radially-polarised  Laguerre-Gaussian modes traversing a magneto-optic crystal lead to a mode propagating within the magneto-optic crystal  with a  polarization texture that can be controlled by an applied magnetic field,  the optical path through the crystal as well as by additional optical elements (such as polarisers) at the output.The same procedure can be followed in the case of an azimuthally-polarised mode, but we have not provided any details for that case.
We have considered two scenarios.  First, when the beam waist $w_0$ is sufficiently large so that the transverse field components dominate the texturing and the longitudinal component is negligible, and, secondly, when the longitudinal component is substantial and so modifies the texturing.   

The coherence of the outgoing optical beam is ensured  by the spin coherence of the magnetic sample (driven to a uniform state by the external magnetic field) and hence structural disorder or impurities in the sample do not alter the obtained photonic polarisation texture. The resulting optical spin angular momentum and intensity textures carry information on the magnetic state (and can thus be used to assess this state) but may also serve further purposes: For instance, from a spectroscopy/microscopy point of view the photonic spin angular momentum structuring allows for sub-wavelength resolution, for excitations of nano or sub-nanoscale objects depends on the orientation of the local polarization. Thus, sensing for these excitations (e.g., via photoelectron spectroscopy) delivers information on the position of the nano object.  Also, polarization texturing allows to access  new types of excitations of nano objects. For instance in electronic systems, azimuthal polarization induces effectively magnetic dipole transitions while radial polarization triggers bulk plasmon excitations \cite{PhysRevB.99.085425}.
The evolving field polarisation distributions shown in  Figure 1 which emerged from an input field that is purely radially-polarised  can be interpreted as superpositions of radial and azimuthal polarizations. Therefore, such fields result in a coherent superposition of magnetic dipole and electric bulk plasmon excitations which falls in the category of chiral transitions. This interpretation is no longer appropriate in presence of a substantial longitudinal component. 
 Although we have focused here on the role of the Faraday effect on radially-polarised light, the Faraday effect is expected to influence the polarisation of other vortex light modes.
The polarization maps conversion in the way discussed here may offer opportunities for magnetic applications. As shown experimentally \cite{PhysRevLett.128.077401}, the magnetization dynamics is sensitive to the orbital and spin angular momentum of the light. A  magnetic or non-magnetic specimen  deposited at the end of our magnetic material discussed above is subject to the output beam which can be modified in situ by tuning the external magnetic field. Changes in the spin and orbital angular momentum as well as the helicity of the output beam of a general vortex light mode (which would involve right and left circular polarisation) can be explored as functions of the magnetic field strength with additional opportunities for fundamental science and applications.\\
\section{Disclosures} The authors declare no conflicts of interest.
\section{Acknowledgments} JB acknowledges financial support through the DFG, project number $429194455$ and TRR$227$ B$06$. 
\bibliography{References}

\end{document}